\documentstyle[proceedings,numreferences,epsf,subfigure,times]{crckapb}
\newcommand{\figwidth}{5.6cm}
\newcommand{\goodgap}{%
	\hspace{\subfigtopskip}%
	\hspace{\subfigbottomskip}}
\newcommand{\beq}{\begin{equation}}
\newcommand{\eeq}{\end{equation}}
\newcommand{\bea}{\begin{eqnarray}}
\newcommand{\eea}{\end{eqnarray}}
\newcommand{\m}{\mu}
\newcommand{\ra}{\rightarrow}
\begin{opening}
\title{Some Pieces of Lattice Evidence in Favor of the
Center-Vortex Picture of Color Confinement$^*$}
\author{M.\ Faber}
\institute{Institut f\"ur Kernphysik, Technische Universit\"at Wien,\\
A--1040 Vienna, Austria}
\author{J.\ Greensite}
\institute{Physics and Astronomy Department, San Francisco State\\
University, San Francisco, CA~94117, USA, and\\
Theory Group, Lawrence Berkeley National Laboratory,\\
Berkeley, CA~94720, USA}
\author{{\v S}.\ Olejn{\'i}k}
\institute{Institute of Physics, Slovak Academy of Sciences,\\
SK--842 28 Bratislava, Slovakia}
\author{D.\ Yamada}
\institute{Physics and Astronomy Department, San Francisco State\\ 
University, San Francisco, CA~94117, USA}
\end{opening}
\runningtitle{Center-Vortex Picture of Color Confinement}
\runningauthor{M.\ Faber et al.}
\begin{document}
%
%
\renewcommand{\thefootnote}{\fnsymbol{footnote}}
\footnotetext[1]{Invited talk presented by {\v S}.\ Olejn{\'\i}k.
His work was supported in part by 
the Slovak Grant Agency for Science (Grant VEGA No.\ 2/4111/97).}
\renewcommand{\thefootnote}{\arabic{footnote}}
%
%
\section{Introduction}
  The center-vortex model of color confinement was proposed more than
20 years ago by `t Hooft~\cite{tHooft:1978}  
and other authors~\cite{others}. It was quite
popular for a while, but then fell in oblivion, before being 
subjected to thorough tests in simulations of quantum chromodynamics 
on a lattice. One of the exceptions was the work of 
Tomboulis and collaborators \cite{Tomboulis:1999a}.
In recent years the interest in this picture has been renewed due
to our discovery of center dominance in SU(2) lattice gauge
theory in maximal center gauge~\cite{DelDebbio:1997a}. 
Substantial lattice evidence has then been 
accumulated on the role played by center vortices in color 
confinement~\cite{DelDebbio:1998b,Langfeld:1998,%
deForcrand:1999}. 

   Our procedure for identifying center vortices is based on
center projection in maximal center gauge. This
gauge brings each link variable as close as possible, 
on average, to a $Z_N$ center element, while preserving a residual 
$Z_N$ gauge invariance. Center projection then is a mapping of 
each SU($N$) link variable to the closest $Z_N$ center element.

   The results that I would like to discuss here try to answer two
questions related to our vortex-identification procedure:

   1.\ How does the procedure work, why is it expected to locate
center vortices relevant for confinement, and why does it in some cases 
fail, on lattice configurations preconditioned in a special way? (See 
Sec.~\ref{VFP} and more details in Ref.~\cite{Faber:1999b}.)

   2.\ A lot of numerical evidence favors the vortex-condensation 
theory in the SU(2) lattice gauge theory, but almost nothing has been
done for SU(3). Does the mechanism work the same way for 3 colors
instead of 2? (See Sec.~\ref{SU3} and Ref.~\cite{Faber:1999c}.)

%
%
\section{Vortex-Finding Property}\label{VFP}
%
%
\subsection{Maximal Center Gauge in SU(2)}
   In SU(2) lattice gauge theory, the maximal center gauge is defined as a
gauge in which the quantity
\beq
       {\cal{R}}[U] = \sum_x \sum_\m \Bigl| \mbox{Tr}[U_\m(x)] \Bigr|^2
\label{e1}
\eeq
reaches a maximum. It is worth noting that this gauge condition is
equivalent to a Landau gauge-fixing condition on adjoint links, i.e.
\beq
      {\cal{R}}[U^A] = \sum_x \sum_\m \mbox{Tr}[U^A_{\m}(x)]
\label{e1adj}
\eeq
should be a maximum. 

   The simplest condition which a method for locating center
vortices has to fulfill is to be able to find vortices inserted 
into a lattice configuration ``by hand''.  
This will be called the {\em ``vortex-finding property''\/}.
Is this condition fulfilled when center vortices are located
by our procedure of center projection in MCG?

   One can argue for an affirmative answer in the following way:
A center vortex is created, in a configuration {$U$}, by making
a discontinuous gauge transformation.  Call the result {$U'$}.  
Apart from the vortex core, the corresponding link variables in
the adjoint representation, {$U^{A}$} and {$U'^A$}, 
are gauge equivalent.  Let {${\cal R}[U^A] = \mbox{max}$} be a
complete gauge-fixing condition (e.g.\ adjoint Landau gauge) on the
adjoint links.  Then (ignoring both Gribov copies and the core region) 
{$U^A$} and {$U'^A$} are mapped into the same gauge-fixed configuration 
{$\tilde{U}^A$}. The original fundamental link configurations
{$U$} and {$U'$} are thus transformed by the gauge-fixing procedure into 
configurations {$\tilde{U},\ \tilde{U}'$} which correspond to the {\em same\/}
{$\tilde{U}^A$}.  This means that {$\tilde{U},\ \tilde{U}'$} can differ only 
by continuous or discontinuous {$Z_2$} gauge transformations, with the
discontinuous transformation corresponding to the inserted center vortex
in {$U'$}. Upon center projection, {$\tilde{U},\ \tilde{U}' 
\rightarrow Z,\ Z'$}, and the projected 
configurations {$Z,\ Z'$} differ by the same discontinuous
{$Z_2$} transformation.  The discontinuity shows up as an additional thin 
center vortex in {$Z'$}, not present in {$Z$}, 
at the location of the vortex inserted by hand.

   However, there are two caveats that could invalidate the argument:

   1.\ We have neglected the vortex core region; and 

   2.\ Fixing to ${\cal R}[U^A]=\mbox{max}$ suffers from the existence
of Gribov copies. 

   To find out whether these caveats are able to ruin 
the vortex-finding property, we have carried out a series of lattice tests. 
The simplest is the following:

   1.\ Take a set of equilibrium SU(2) configurations. 

   2.\ From each configuration make three:

   I -- the original one;

   II -- the original one with 
{$U_4(x,y,z,t)\rightarrow (-\mbox{\bfseries1})\times U_4(x,y,z,t)$} 
for {$t=t_0$},
{$x_1\le x\le x_2$} and all {$y$, $z$},
i.e.\ with {2 vortices (one lattice spacing thick) inserted by hand}.
Moreover, a random (but continuous) gauge transformation 
is performed on the configuration with inserted vortices;

   III -- a random copy of {I}.

   3.\ Measure:
\beq
G(x)=\frac{\sum_{y,z} <P_{I}(x,y,z) P_{II}(x,y,z)>}
     {\sum_{y,z} <P_{I}(x,y,z) P_{III}(x,y,z)>}\;.
\eeq
{$P_i(x,y,z)$} is the Polyakov line measured on the 
configuration {$i={}$I, II, or III}.

   If the method correctly identifies the inserted vortices, one expects
\beq
       G(x) = \left\{ \begin{array}{rl}
            -1 & x \in [x_1,x_2] \cr
             1 & \mbox{otherwise} \end{array} \right.\;.
\label{Gexp}
\eeq
%

%
%
\begin{figure}[t!]
\centering
\subfigure[Configurations fixed directly to the maximal center gauge.]%
{\label{fig1-a}\epsfxsize=\figwidth\epsffile{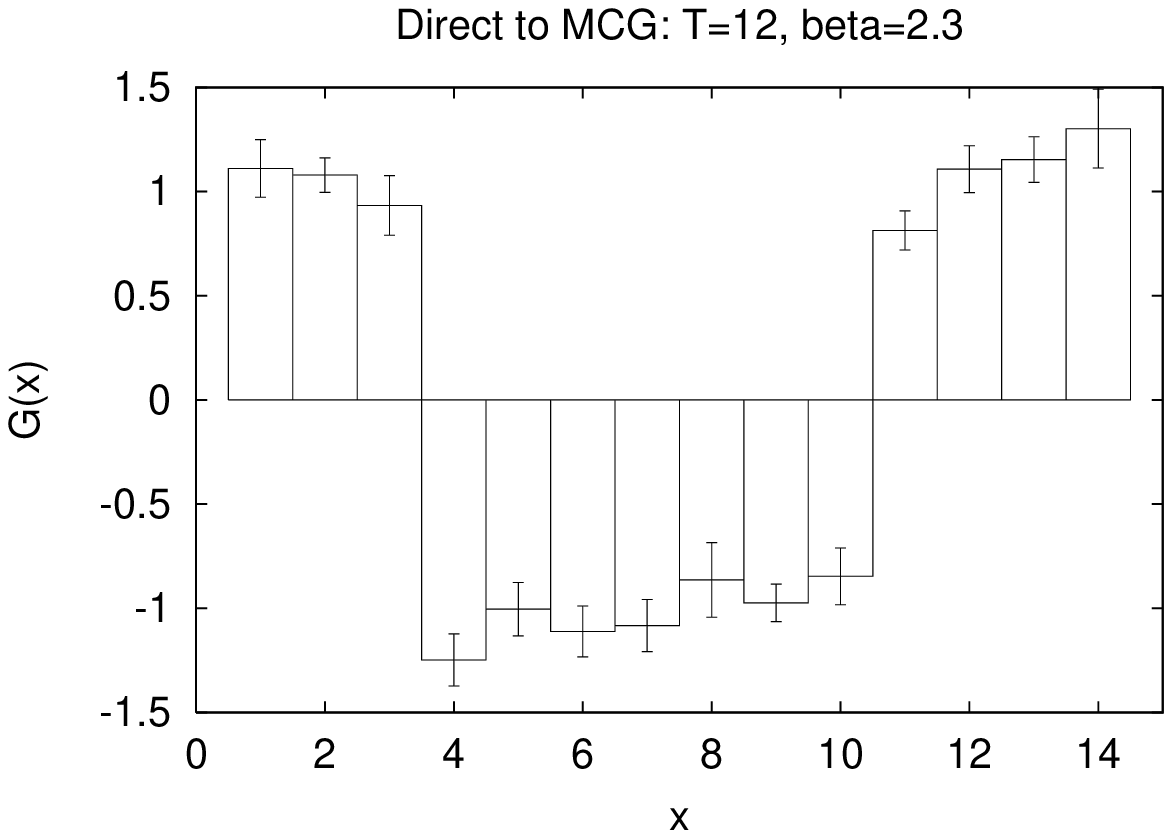}}\goodgap
\subfigure[Configurations first fixed to the Landau gauge, 
then to MCG.]{\label{fig1-b}\epsfxsize=\figwidth\epsffile{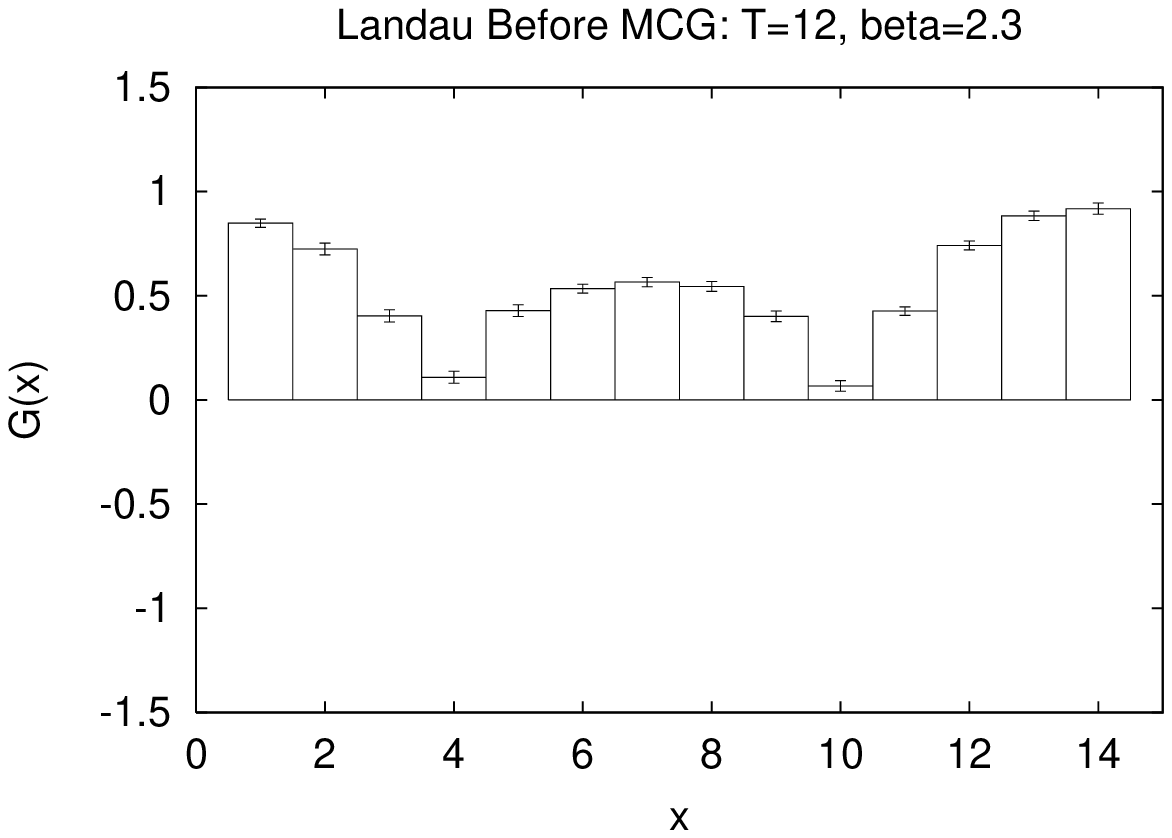}}
\caption{Graph of $G(x)$ for configurations
 with thin inserted vortices ($14^3\times12$ lattice). 
 The discontinuity
 was inserted to the time links within the volume $4\!\leq\!x\!\leq\!10,\; 
 0\!\leq\!y\!<\!14,\; 0\!\leq\!z\!<\!14$ at the time slice $t=T$.}
\label{fig1}
\end{figure}

   The result of the test is shown in Fig. \ref{fig1-a}. The inserted 
vortices are clearly recognized, and the associated Dirac volume
is found in its correct location. The vortex-finding property is also
preserved when we insert vortices with a core a few 
lattice spacings thick. 

%
%
\subsection{MCG Preconditioned with Landau Gauge}
   It has been shown recently by Kov\'acs and 
Tomboulis~\cite{Kovacs:1999b} that if one first fixes to 
Landau gauge (LG), before relaxation to maximal center gauge, center 
dominance is lost. They argued that this problem casts some doubt on the
physicality of objects defined through center projection in MCG. 

   This failure has a simple explanation: {\em LG preconditioning 
destroys the vortex-finding property\/}.  This is illustrated by redoing 
the test shown in Fig.\ \ref{fig1-a}, only with a prior fixing to 
Landau gauge.  The result, shown in Fig.\ \ref{fig1-b}, is that the
vortex-finding condition is not satisfied; the Dirac
volume is not reliably identified.

   The Gribov copy problem, which is fairly harmless on most of the
gauge orbit~\cite{DelDebbio:1998b}, seems severe enough to 
ruin vortex-finding on a tiny region of the gauge orbit near Landau gauge.

%
%
\subsection{Other Gauges}
   The vortex-finding argument above does not seem to single out MCG.
In fact, there should exist (infinitely) many gauges with the vortex-finding
property. They should fulfill the following
requirements:

   1.\ The gauge condition depends only on the {\em adjoint\/}
representation links.

   2.\ It is a complete gauge-fixing for adjoint link variables.

   3.\ The gauge fixing transforms most links to be close to center elements,
at weak coupling.

   We have tried a couple of gauge conditions, and the results are summarized
in Table~1. It turns out in all cases that the loss of
the vortex-finding property goes hand in hand with the loss of center 
dominance. A notable fact is that the recently proposed 
Laplacian center gauge~\cite{Alexandrou:1999}, 
which is free of the Gribov-copy problem,
is a perfect (thin-)vortex finder, i.e.\ the simple expectation
(\ref{Gexp}) is satisfied by the numerical data \emph{exactly}.

%
%
\begin{table}[htb]
\begin{center}
\label{tab1}
\caption{{Comparison of various adjoint gauges. (Details see in 
Ref.~\protect\cite{Faber:1999c}.)}}
\begin{tabular}{lcc}
\hline
{\bf Gauge}&{\bf Center dominance}&{\bf Vortex-finding property}\\
\hline
{maximal center gauge}&{yes}&{yes}\\
{Landau gauge, then MCG}&{no}&{no}\\
{asymmetric adjoint gauge\protect%
\footnote{}
}&{yes}&{yes}\\
{adjoint Coulomb gauge}&{no}&{no}\\
{``modulus'' Landau gauge}&{no}&{no}\\
{MCG, then ``modulus'' Landau gauge}&{yes}&{yes}\\
{Laplacian center gauge}&{yes}&
{{perfect!}}\\
\hline
\end{tabular}
\end{center}
\end{table}

%
%
\section{Evidence for Center Dominance in SU(3) Lattice Gauge Theory}
\label{SU3}
%
%
\subsection{Maximal Center Gauge in SU(3)}
   The maximal center gauge in SU(3) gauge theory is defined as the gauge
which brings link variables $U$ as close as possible to elements of
its center $Z_3$. This can be achieved as in SU(2) by maximizing 
a ``mesonic'' quantity
\beq\label{mesonlike}
       {\cal{R}} = \sum_x \sum_\m \Bigl| \mbox{Tr}[U_\m(x)] \Bigr|^2,
\eeq
or, alternatively, a ``baryonic'' one
\beq\label{baryonlike}
{\cal{R}}'=
\sum_x\sum_\mu\mbox{Re}\left(\left[\mbox{Tr}\;U_\mu(x)\right]^3\right).
\eeq
The latter was the choice of Ref.\ \cite{DelDebbio:1998b}, 
where we used the method
of simulated annealing for iterative maximization procedure. The 
convergence to the maximum was rather slow and forced us to restrict
simulations to small lattices and strong couplings.

%
%
\footnotetext[\thefootnote]%
{This is a slight generalization of MCG, namely a gauge
maximizing the quantity
${\cal R}'[U]={\sum_{x,\mu}}\;c_\mu\;| \mbox{Tr}[U_\mu(x)] |^2\;$
with some choice of {$c_\mu$}, e.g.\ $c_\mu=\{1,1.5,0.75,1\}$.}

   The results, that will be presented below, were obtained in a gauge 
defined by the ``mesonic'' condition (\ref{mesonlike}). The maximization
procedure for this case is inspired by the Cabibbo--Marinari--Okawa
SU(3) heat bath method~\cite{Cabibbo:1982}.
The idea of the method is as follows: In the maximization
procedure we update link variables to locally maximize the quantity
(\ref{mesonlike}) with respect to a chosen link. 
Instead of trying to find the optimal gauge-transformation 
matrix $\Omega(x)$, we take an 
SU(2) matrix $g(x)$ and embed it into one of the three 
diagonal SU(2) subgroups of SU(3). The expression 
for a chosen link is then maximized with respect to $g$, 
with the constraint of $g$ being an SU(2) matrix.
This reduces to an algebraic problem (plus a solution of a non-linear
equation). Once we obtain the matrix $g(x)$, we update link variables
touching the site $x$, and repeat the procedure for all three subgroups
of SU(3) and for all lattice sites. This constitutes one center
gauge fixing sweep.
Center projection is then done by replacing each link matrix by the
closest element of $Z_3$.

   The above iterative procedure was independently developed by
Montero. Its detailed description is contained in his recent publication
\cite{Montero:1999}. 

%
%
\subsection{Center Dominance in SU(3) Lattice Gauge Theory}
   The effect of creating a center vortex linked to a given Wilson loop
in SU(3) lattice gauge theory is to multiply the Wilson loop by an element
of the gauge group center, i.e.\ $W(C) \ra e^{\pm2\pi i/3} W(C)$.
Quantum fluctuations in the number of vortices linked to a Wilson loop
can be shown to lead to its area law falloff; the simplest, but 
urgent question is whether center disorder is sufficient to produce
the whole asymptotic string tension of full, unprojected lattice 
configurations.

\begin{figure}
\centering
\subfigure[Creutz ratios for the original, the $Z_3$ projected, 
and the modified (with vortices removed) ensembles. 
The value of $\chi(4)$ (shown in grey) comes from the compilation of 
Levi~\protect\cite{Levi}.]{\label{chi_vs_R}%
\epsfxsize=\figwidth\epsffile{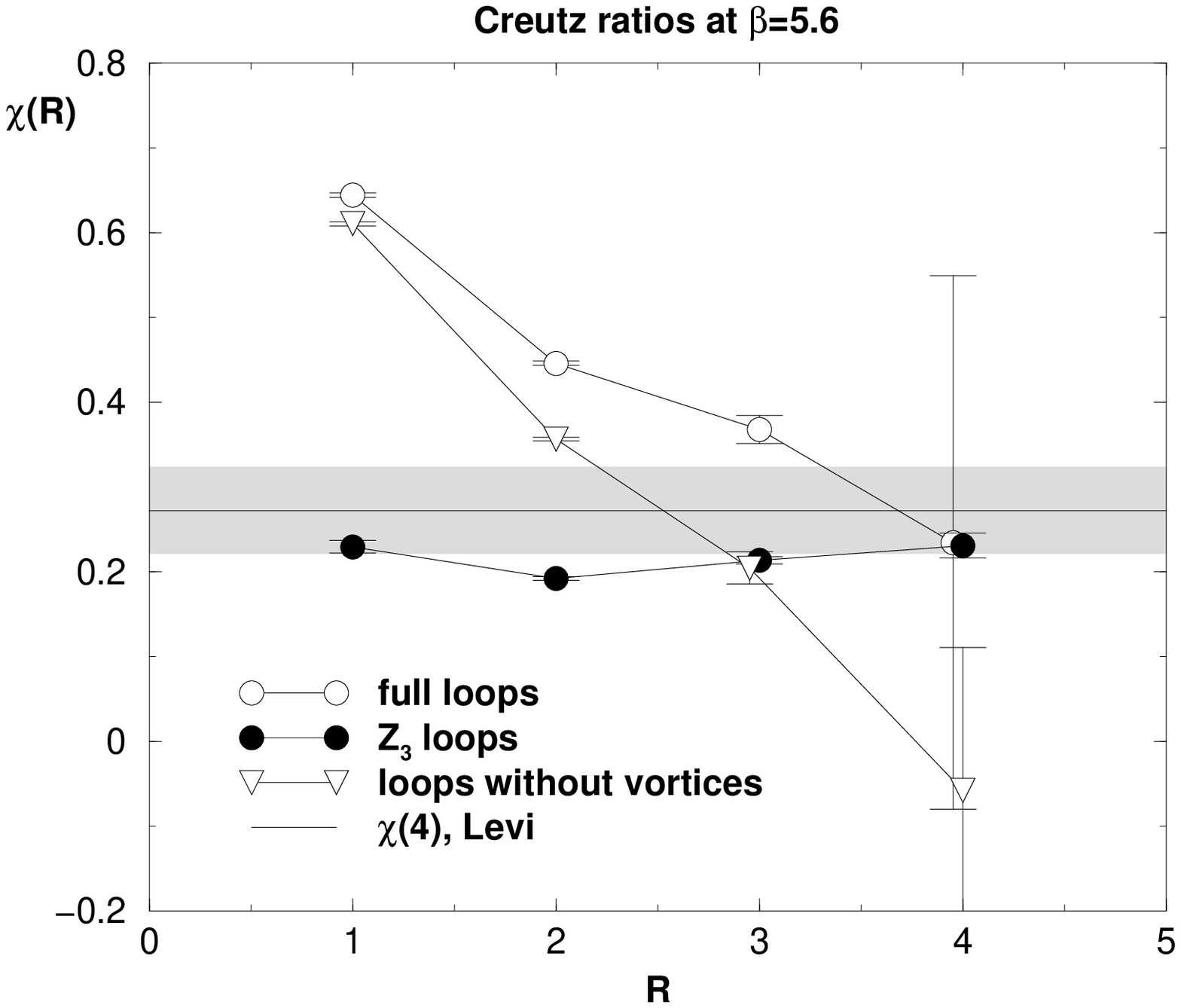}}\goodgap
\subfigure[Center-projected Creutz ratios vs.\ $\beta$. Full
circles connected with a solid line are asymptotic values 
quoted by Bali and Schilling~\protect\cite{Bali:1993}.]{\label{chi_vs_b_with_bali}%
\epsfxsize=\figwidth\epsffile{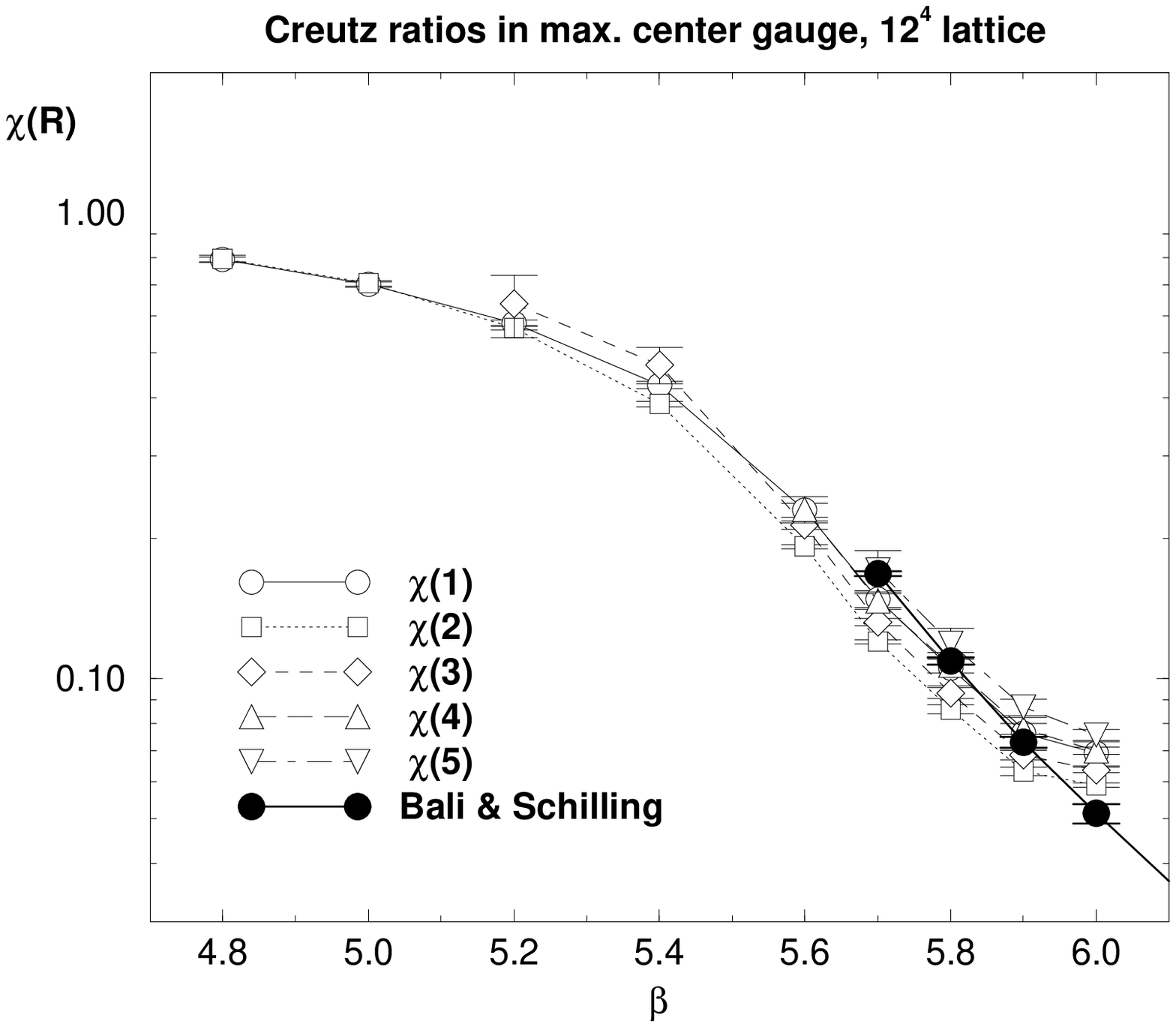}}
\caption{Creutz ratios in SU(3) lattice gauge theory.}
\label{fig2}
\end{figure}

   We have computed Wilson loops and Creutz ratios at various values
of the coupling $\beta$ on a $12^4$ lattice, from full lattice
configurations, center-projected link configurations in
maximal center gauge, and also
from configurations with all vortices removed. Figure \ref{chi_vs_R}
shows a typical plot at $\beta=5.6$. It is obvious that center
elements themselves produce a value of the string tension which
is close to the asymptotic value of the full theory. On 
the other hand, if center elements are factored out from link matrices
and Wilson loops are computed from SU(3)/$Z_3$ elements only, 
the Creutz ratios tend to zero for sufficiently large loops.
The errorbars are, however, rather large, and one cannot
draw an unambiguous conclusion from the data.

%
%

   Center dominance by itself does
not prove the role of center degrees of freedom in QCD 
dynamics~\cite{Ambjorn:1998,Faber:1999a};
some sort of center dominance exists also without any gauge fixing
and can hardly by attributed to center vortices. Distinctive
features of center-projected configurations in
maximal center gauge in SU(2), besides center dominance, were that: 

   1.\ Creutz ratios were approximately constant starting from
small distances (this we called ``precocious linearity''),

   2.\ the vortex density scaled with $\beta$ exactly as expected
for a physical quantity with dimensions of inverse area.
 
   Precocious linearity, the absence of the Coulomb part of the
potential on the center-projected lattice at short distances,
can be quite clearly seen from Fig.\ \ref{chi_vs_R}. One observes
some decrease of the Creutz ratios at intermediate distances. A similar effect
is present also at other values of $\beta$. It is not clear to us
whether this decrease is of any physical relevance, or whether it should be
attributed to imperfect fixing to the maximal center gauge.

   The issue of scaling is addressed in Figure \ref{chi_vs_b_with_bali}.
Here values of various Creutz ratios are shown as a function of $\beta$
and compared to those quoted in Ref.\ \cite{Bali:1993}. All values
for a given $\beta$ lie close to each other and are in reasonable 
agreement with asymptotic values obtained
in time-consuming SU(3) pure gauge theory simulations. The plot in 
Fig.\ \ref{chi_vs_b_with_bali} is at the same time a hint
that the P-vortex density also scales properly. 
The density is approximately proportional 
to the value of $\chi(1)$ in center-projected configurations, and 
$\chi(1)$ follows the same scaling curve as Creutz ratios obtained
from larger Wilson loops. 

   A closer look at Fig.\  \ref{chi_vs_b_with_bali} reveals that
there is no perfect scaling, similar to the SU(2) case, in our SU(3)
data. Broken lines connecting the data points tend to bend
at higher values of $\beta$. In our opinion, this is a finite-volume
effect and should disappear for larger lattices. 

%
%
\section{Conclusions}
   I conclude with the following comments/conclusions:

   1.\ Center vortices are created by discontinuous gauge transformations,
which make no reference to any particular gauge condition. 
However, appropriate gauge fixing is necessary 
to reveal relevant center vortices:
In MCG -- and in an infinite class of other adjoint gauges -- such 
discontinuous transformations are squeezed to the identity everywhere
except on Dirac volumes, whose locations (together with those of
the associated vortices) are then revealed upon center projection.

   2.\ If the vortex-finding property is destroyed by some modification
of the gauge-fixing and center-projection procedure, then 
center vortices are not correctly identified on thermalized lattices, 
and center dominance in the projected configurations is lost.
This fact does not call
into question the physical relevance of P-vortices found by our usual
procedure (which \emph{has} the vortex-finding property); that 
relevance is implied by the strong correlation that was shown to exist
between these objects and gauge-invariant observables.

   3.\ Center dominance is quite clearly seen also in SU(3) lattice gauge 
theory, however, more convincing data and better gauge-fixing procedure 
is needed.

%
%
\noindent{\bf Acknowledgements}\\[1mm]
I am grateful to the organizers of this excellent Workshop, 
especially Valja Mitrjushkin, for invitation and warm hospitality.

%
%

\end{document}